\documentclass[%
reprint,
superscriptaddress,
nofootinbib,
 amsmath,amssymb,
 aps,
 prapplied,
floatfix,
floats,
raggedbottom,
longbibliography,
]{revtex4-1}

\usepackage{float}
\usepackage{appendix}

\usepackage{physics}
\usepackage{mathtools}
\usepackage{graphicx}
\usepackage{dcolumn}
\usepackage{bm}

\usepackage{xcolor}
\usepackage{soul}

\AtBeginDocument{%
    \newwrite\bibnotes
    \def\bibnotesext{Notes.bib}
    \immediate\openout\bibnotes=\jobname\bibnotesext
    \immediate\write\bibnotes{@CONTROL{REVTEX41Control}}
    \immediate\write\bibnotes{@CONTROL{%
    apsrev41Control,author="08",editor="1",pages="1",title="0",year="1"}}
     \if@filesw
     \immediate\write\@auxout{\string\citation{apsrev41Control}}%
    \fi
}%

\begin{document}

\newcommand{\EZcomment}[1]{{\color{red}{EZ:#1}}}
\newcommand{\EZcrossout}[1]{{\st{EZ:#1}}}
\newcommand{\AGsug}[1]{{\color{teal}{#1}}}
\newcommand{\AGcomment}[1]{{\color{teal}{AG:#1}}}
\newcommand{\AGcrossout}[1]{{\st{AG:#1}}}
\newcommand{\LQcomment}[1]{{\color{violet}{LQ:#1}}}
\newcommand{\LQcrossout}[1]{{\st{LQ:#1}}}
\newcommand{\BKcomment}[1]{{\color{orange}{BK:#1}}}
\newcommand{\BKcrossout}[1]{{\st{BK:#1}}}

\preprint{APS/123-QED}

\title{Machine-Learning-Derived Entanglement Witnesses}

\author{Alexander C. B. Greenwood}
\thanks{alexander.greenwood@mail.utoronto.ca; A.C.B.G. and L.T.H.W. contributed equally to this work.}

\author{Larry T. H. Wu}
\thanks{alexander.greenwood@mail.utoronto.ca; A.C.B.G. and L.T.H.W. contributed equally to this work.}
\author{Eric Y. Zhu}
\affiliation{%
Dept of Electrical \& Computer Engineering, University of Toronto \\
Toronto, Ontario, Canada M5S 3G4
}%
\author{Brian T. Kirby}
\affiliation{Tulane University, New Orleans, LA 70118 USA}
\affiliation{DEVCOM Army Research Laboratory, Adelphi, MD 20783 USA}
\author{Li Qian}
\affiliation{%
Dept of Electrical \& Computer Engineering, University of Toronto \\
Toronto, Ontario, Canada M5S 3G4
}%

\date{\today}

\begin{abstract}
In this work, we show a correspondence between linear support vector machines (SVMs) and entanglement witnesses, and use this correspondence to generate entanglement witnesses for bipartite and tripartite qubit (and qudit) target entangled states.
An SVM allows for the construction of a hyperplane that clearly delineates between separable states and the target entangled state; this hyperplane is  a weighted sum of observables (‘features’) whose coefficients are optimized during the training of the SVM. 
We demonstrate with this method the ability to obtain witnesses that require only local measurements even when the target state is a non-stabilizer state. 
Furthermore, we show that SVMs are flexible enough to allow us to rank features, and to reduce the number of features systematically while bounding the inference error. This allows us to derive $W$ state witnesses capable of detecting entanglement with fewer measurement terms than the fidelity method dominant in today's literature.
The utility of this approach is demonstrated on quantum hardware furnished through the 
IBM Quantum Experience.

\end{abstract}

\maketitle


\section{Introduction}
The entanglement of high-dimensional quantum systems is the critical enabling resource in many applications of quantum information science, quantum communications \cite{ali2007large, zhong2015photon}, imaging \cite{chen2014quantum}, and information processing \cite{lu2019quantum}.
The systems with the smallest dimension capable of exhibiting entanglement are those of two qubits.  Two possible routes to realizing quantum systems with higher dimensionality include increasing the number of subsystems (i.e., moving from systems of two qubits to those of \textit{N} qubits) \cite{Thomas2022} or increasing the dimension of the existing subsystems (from $d=2$ qubits to $d>2$ to qudits) \cite{kues2017chip, imany2019high}. 
Current developments in quantum technologies are adopting both approaches.

It is therefore crucial to have an 
efficient method that allows us to experimentally detect the presence of entanglement in high-dimensional quantum systems. 
The brute force approach is to fully characterize a system by performing quantum state tomography and calculating separability 
measures from the recovered density matrix. However, tomography is experimentally and computationally demanding.
For a state consisting of $N$ particles, with each residing in a d-dimensional Hilbert space, we would have to perform $M = 
O(d^{2N})$ measurements \cite{PhysRevA.66.012303}. In addition to the sheer number of measurements required, there is also the computational cost 
of regression to recover the density matrix. 
With full characterization of a high-dimensional quantum state being so expensive experimentally and computationally, classification of a state becomes a more attractive option.  

Recent studies of the entangled-separable quantum state classifiers have utilized aspects of machine learning, such as neural networks 
\cite{ma2018transforming_ML_Entangle}, and convex hull approximations \cite{lu2018separability}.  
Yet others have attempted to improve resources scaling of full state tomography by replacing or augmenting conventional approaches to state reconstruction (such as those based on maximum-likelihood estimation \cite{DFVJames_2001} or Bayesian methods \cite{Lukens_2020}) using deep learning methods \cite{lohani2020machine,Danaci_2021,lohani2022demonstration}.
However, the number of features (or observables) required to provide correct classification and estimation for such systems often grow to the number required for full state tomography.

A more efficient approach is to construct an observable known as an entanglement witness $\hat{W}$.  
The expectation value of this observable would give a non-negative value for all separable states, while a target entangled state would give a negative value. Simply 
measuring the expectation value $\langle\hat{W}\rangle$ for a given system is enough to tell us if it is close to the desired entangled state, 
without the need to find the full density matrix. There is also at least a quadratic reduction in the number of measurements required, $O(d^N)$, compared to $O(d^{2N})$ for state tomography 
\footnote{For an N-particle d-dimensional space, we posit that $\hat{W}$ will have no more than $(d+1)^N$ features; we arrive at this expression because a particle of dimension $d$ will have at most $(d+1)$ mutually-unbiased bases, with each basis corresponding to its own generalized Pauli matrix $\sigma_k$.}. 

There is extensive literature on constructing entanglement witnesses.  These include using the stabilizer formalism \cite{toth2005entanglement_stabilizer} to derive witnesses for multi-partite, graph and cluster states. 
Witnesses can also be constructed through the ‘fidelity method’ \cite{bourennane2004experimental}:
\begin{equation}
\hat{W}=cI-|\psi_{\mathrm{target}}\rangle \langle \psi_{\mathrm{target}} |, 
\end{equation}
with $c$ simply a classical c-number, and $|\psi_{\mathrm{target}}\rangle $  the target state. However, this often results in witnesses that require a large number of measurements \cite{bourennane2004experimental}.

\begin{figure*}
    \includegraphics[width=14cm]{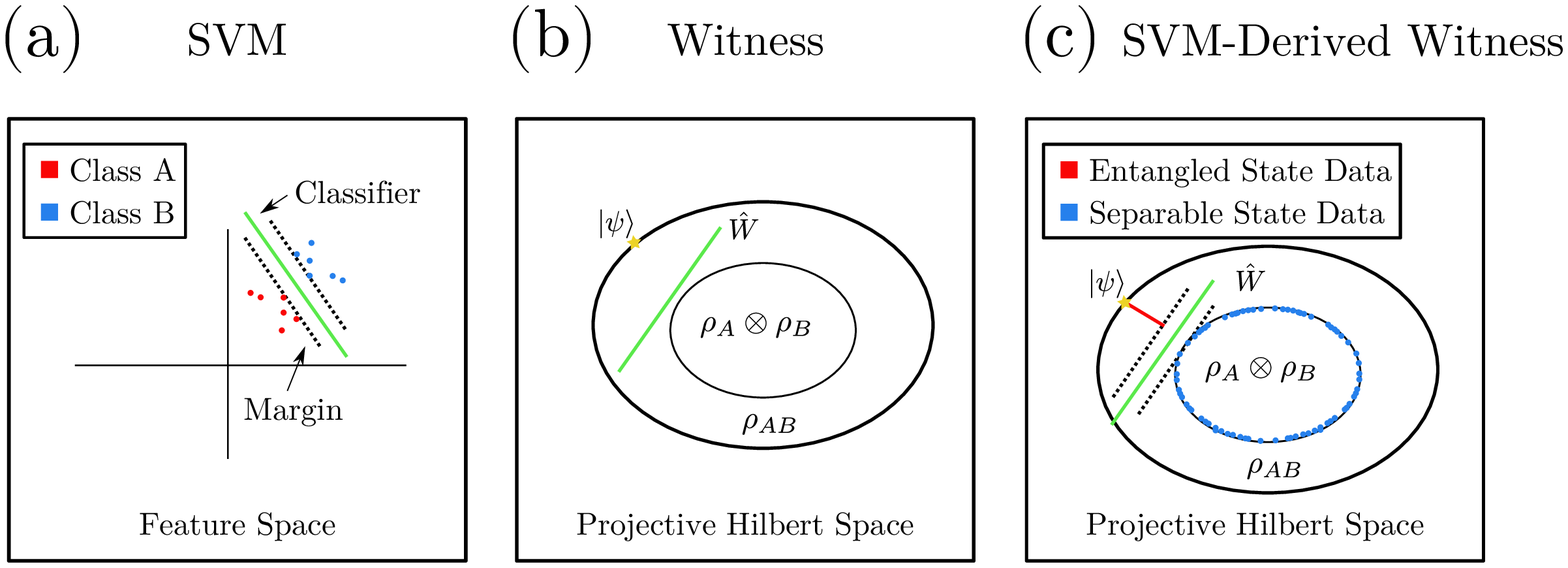}
    \caption{(a) An illustration of a Support Vector Machine (SVM) in feature space for a bipartite system. A hyperplane (green solid line) defined by an SVM is separated from two classes of data A and B by the maximally allowed margin (dashed black line). (b) An entanglement witness $\hat{W}$ (green solid line) can be viewed as a hyperplane in \textit{Projective Hilbert Space} that separates all separable states (which form a convex set) from a subset of entangled states. (c) A depiction of an SVM-derived entanglement witness that combines the concepts from (a) and (b). As shown in Section I, entangled state data is sampled from a set of Werner states, which can be viewed as a line segment between a target state $\ket{\psi}$ and the maximally mixed state at the center of the separable state region. The length of the line segment is determined by the range of p in Eq(6): the larger the p, the closer is the entangled Werner state to the separable states. All separable states used for training throughout this work are pure and lie along the surface of the set of separable states.}
    \label{fig:summary_of_svmdew}
\end{figure*} 
In this work, we note the analogy between linear support vector machines (SVMs) \cite{vapnik1963pattern} and entanglement witnesses and use this as a means of deriving  witnesses for entangled states.  
An SVM is a supervised machine learning (ML) technique that uses a hyperplane to perform binary classification.  This is analogous to an entanglement witness $W$; on one side of the hyperplane lie all the separable states ($tr(\rho \hat{W})\geq 0$), while the other side ($tr(\rho \hat{W})< 0$) contains only entangled states, including the target state. A depiction of the analogy between SVMs and entanglement witnesses is shown in Fig. \ref{fig:summary_of_svmdew}.

We will limit ourselves to local measurements; the ansatz we will use for the witness is:
\begin{equation}
	\hat{W} = \sum_{k_1, k_2, \cdots, k_N}  a_{k_1, k_2, \cdots, k_N} 
	\sigma_{k_1} \otimes \sigma_{k_2} \otimes \cdots \otimes \sigma_{k_N}.
	\label{eqn:W_expect}
\end{equation} 
\noindent
Each term of the witness is simply a string of (generalized) Pauli matrices $\sigma_k \in \{I,X, Y, Z\}$ with a real weight $a_{k}$. The length of the string corresponds to the number $N$ of qubits (qudits) in the system.  
In the case of qudits ($d>2$), each Pauli string would be a sum of the generalized Pauli string and its Hermitian conjugate. More details can be found in Section III.

The training of the SVM consists of first generating the separable and entangled states.  
For each of these states, the expectation values of all the Pauli strings 
($x_{\vec{k}}\equiv  \langle \sigma_{k_1} \otimes \sigma_{k_2} \otimes \cdots \otimes \sigma_{k_N} \rangle $) are then computed; these will be the `features' used to train the SVM.  Optimization of the SVM involves varying the coefficients $a_{\vec{k}}$ so that the weighted sum of the features 
($\langle\hat{W}\rangle = \sum_{\vec{k}} a_{\vec{k}} x_{\vec{k}}$) 
gives the correct classification for each state.  
We note that, unlike many other ML techniques such as deep learning, the training of an SVM is convex, meaning if a solution exists for the given target state and ansatz, the \emph{optimal} SVM \footnote{By `optimal', we mean that the hyperplane is ‘optimally’ trained with respect to the training data and ansatz, not that the resulting witness is ‘optimal’.} will be found.   

In what follows, we will demonstrate that our SVM approach allows for the derivation of entanglement witnesses for bipartite (Section I) and tripartite qubit states (Section II), as well as bipartite qudit states (Section III), and show how the technique can be extended to higher dimensions and particles. 
In particular, we recover a witness for tripartite $W$ states that utilizes only local measurements, but requires far fewer features and has comparable noise tolerance to what has previously been demonstrated \cite{guhne2003investigating,bourennane2004experimental}.
Additionally, we show that our SVM formalism allows for the programmatic removal of features, i.e., 
reducing the number of experimental measurements, in exchange for a lower tolerance to white noise, in a manner similar to \cite{Sciara_PhysRevLett.122.120501}.

In Section IV, we verify our derived witnesses for the tripartite W-state on a physical system in the form of a quantum circuit run on the IBMQ cloud and compare its performance to the standard fidelity witness (Eq 1) using the same platform. We find the derived witnesses to have comparable immunity to decoherence than the fidelity-based witness. Perhaps the most salient feature, we show that a SVM-derived witness implemented on real IBMQ hardware requires fewer measurements to verify entanglement than the fidelity-method counterpart.

The use of statistical learning methods for deriving witnesses is a largely unexplored area. This work calls into question the criteria by which an operator could be practically proven as a formal entanglement witness. We shall leave any rigorous analysis and optimization of our witnesses for future work; the work that follows merely serves as a proof-of-concept for a potentially powerful technique capable of characterizing entanglement in high-dimensional systems.

\section{Basic Scheme and Application to a  Simple Example}

As we saw in Eqn \ref{eqn:W_expect} and the discussion that followed, an SVM is a linear classifier that takes a weighted sum of 
`features' (expectation values $x_{\vec{k}}$ ) of an object (quantum state) and predicts whether it belongs in one class (entangled) or another (separable).  More formally, we can write the classifier as:
\begin{equation}
	\begin{array}{l}
	y = \sum a_{\vec{k}} x_{\vec{k}}  \\
	\mathrm{If\ separable: \ } y\geq 0 \\
	\mathrm{If \ entangled: \ }	y< 0. 
	\end{array}
	\label{eq:classifier_constraints}
\end{equation}
What is shown in \eqref{eq:classifier_constraints} refers to a constraint imposed on our \textit{classifier} and is not true for entanglement witnesses in general. The coefficients $a_{\vec{k}}$ are `learned' by using a set of training examples, each with label $\hat{y}$ and features $x_{\vec{k}}$; the labels for entangled states will be $\hat{y} = -1$, and $\hat{y} = +1$ for separable states.  
The learning involves the minimization of the loss function:
\begin{equation}
		\mathcal{L} =\frac{1}{T}\sum_{t=1}^T \left[\mathrm{max}(0,1- \hat{y}^{(t)} \cdot y^{(t)})\right]^{m}+\lambda\sum_{\vec{k}} |a_{\vec{k}}|,
\label{eqn:lossfunc}
\end{equation}
with respect to the coefficients $a_{\vec{k}}$.  
The first summation is over all training examples 
 $T$, $\hat{y}^{(t)}$ (${y}^{(t)}$) is the label (prediction) for a particular
example, $m = 1$ or 2, and the second summation (over $\vec{k}$) is a regularization term (whose relative importance can be varied with a scaling factor, $\lambda$) to limit the number of non-zero features.  

The first term of the loss function (Eqn \ref{eqn:lossfunc}) is known as the hinge loss.  
When $m$ is equal to either 1 or 2, the loss function is a convex function of the coefficients $a_{\vec{k}}$; that is, 
the loss function has only a single global minimum for a set of training examples, though the trained coefficients for $m = 1$ will be different from $m=2$.   
We will use  $m=1$ so that 
there are fewer non-zero features present (the SVM uses `sparse' features); however, $m = 2$ may be needed when the separation between the separable training states and the entangled training states is narrow \cite{lee2013study}. Throughout this work, we consider the detection of entangled states that are well separated from the set of separable states. An example in which the two classes of data are close to one another would be the detection of highly mixed entangled states, which lie close to the boundary between the set of entangled and separable states.

To illustrate the training of the SVM, we use a simple illustrative example.  Let us find the entanglement witness for the 
bipartite qubit Bell-state:
\begin{equation}
|\psi_{\mathrm{target}}\rangle = \frac{1}{\sqrt{2}}
   	   \left(  |0 0\rangle +  |1 1\rangle    \right),
\end{equation}
with $|0\rangle$ and $|1\rangle$ being the eigenstates of the Pauli matrix $\sigma_{z}$.
We can construct a set of density matrices (representing the class of entangled states) by adding white noise to a target bell state $\ket{\psi_{target}}$
\begin{equation}
	\rho_{\mathrm{ent}}(p) = (1-p)|\psi_{\mathrm{target}}\rangle\langle\psi_{\mathrm{target}}| + \frac{p}{4}\mathcal{I},
	\label{eq:werner}
\end{equation}
and varying $p$ uniformly over the range $\left[0, \frac{2}{3}\right)$. One could show that \eqref{eq:werner} represents a convex set by virtue of the intermediate points all being convex sums of the states lying at the boundaries defined by $p = 0$ and $ p= 2/3$.  
Each training example for the entangled class will consist of the label $\hat{y}(p) = -1$ and the features
$\{x_{\vec{k}}(p)\} = \{ tr(\rho_{ent}(p) \sigma_{k_1}\otimes \sigma_{k_2})  \}$.

On the other hand, when constructing the  training data for bipartite separable states we can simply use pure states, due to the linear nature of the entanglement witness \footnote{Consider an arbitrary separable state $\rho_{sep} = \sum_i p_i |\psi_{sep,i}\rangle \langle \psi_{sep,i}|$, 
where $\psi_{sep,i}$ are fully-separable states.  Then $tr(\rho_{sep}\hat{W}) = \sum_i p_i \langle \psi_{sep,i}|\hat{W}|\psi_{sep,i}\rangle\geq 0 $. This means that using pure fully-separable states as training examples is equivalent to training with mixed fully-separable states.}.
Random separable states are constructed by taking the tensor product of two haar-distributed, single-qubit states. Each training example $|\psi_{sep}\rangle $ for the separable class consists of the label $\hat{y} = +1$ and features
$\{x_{\vec{k}}\} = \{ \langle{\psi_{sep}}|\sigma_{k_1}\otimes \sigma_{k_2}) |{\psi_{sep}}\rangle \}$.

Four thousand examples of each class are generated using \texttt{NumPy}, a linear algebra library in \texttt{Python}, and the ML library \texttt{Tensorflow} \cite{tensorflow2015-whitepaper} is used  to train the SVM.  
The process is repeated for the three other Bell states, and the results are tabulated in Table \ref{tab:bellstates}.  
Only features in the trained SVM with coefficients whose absolute value greater than 0.01 are shown; 
all other features are discarded. The choice of keeping terms that satisfy this criteria allows us to ignore terms that account for changes in $\abs{\expval{W}}$ of less than $1\%$ of the maximum achievable value.
The witness is verified using $10^4$ samples of separable states and $10^4$ samples of entangled states with noise in the range of $p\in[0, \frac{1}{4}]$ \footnote{One could consider an even larger interval of $p$ over which the entangled states are defined, but with the modification of using an $m = 2$ hinge loss function to account for the decreased separation between sampled entangled and separable states.}. The SVM is observed to give the right classification 
100\% of the time. 

The accuracy of the witnesses in Table 1 is unsurprising; these operators are in exact correspondence with those derived according to Terhal's method prescribed in \cite{TERHAL2002313}. The most remarkable feature is that the theoretically optimal witnesses have been derived from a relatively small data set of $10^4$ points in each category (separable and non-separable).

\begin{table}
\begin{center}
\begin{tabular}{ c | c}
Target State &  Witness \\ \hline
$|\Phi^{+}\rangle = \frac{1}{\sqrt{2}}\left( |00\rangle +|11\rangle \right)$ &  
$ I\otimes {I} - X\otimes {X} + Y\otimes{Y} - Z\otimes{Z}$ \\   \\
$|\Phi^{-}\rangle = \frac{1}{\sqrt{2}}\left( |00\rangle - |11\rangle \right)$ &  
$ I\otimes {I} + X\otimes {X} - Y\otimes{Y} - Z\otimes{Z}$ \\   \\
$|\Psi^{+}\rangle = \frac{1}{\sqrt{2}}\left( |01\rangle + |10\rangle \right)$ &  
$ I\otimes {I} - X\otimes {X} - Y\otimes{Y} + Z\otimes{Z}$ \\ \\
$|\Psi^{-}\rangle = \frac{1}{\sqrt{2}}\left(  |01\rangle - |10\rangle \right)$ &  
$ I\otimes {I} + X\otimes {X} + Y\otimes{Y} + Z\otimes{Z}$ \\
\end{tabular}
\end{center}
\caption{Bell states and their corresponding witnesses \label{tab:bellstates} obtained from our method}
\end{table}

\section{Application of the Method to Tripartite Qubit States}

In this section, we will extend the method presented in Section I to the tripartite qubit states: 
\begin{equation}
\begin{split}
|{GHZ}\rangle & = \frac{1}{\sqrt{2}}\left( |000\rangle+ |111\rangle  \right), \\
|{W}\rangle & = \frac{1}{\sqrt{3}}\left( |001\rangle+ |010\rangle   + |100\rangle \right)
\end{split}
\label{eq:GHZ_W}
\end{equation}

We first point out that, unlike the bipartite Bell states, the GHZ and $W$ states are not interconvertible with 
the use of local unitary operations; they belong to separate classes of entanglement \cite{PhysRevA.62.062314}.  Moreover, \emph{separable} states in tripartite systems can be further divided into (1) fully-separable ($fs$) states:
\begin{equation}
	\rho^{(1,2,3)}_{fs} = \rho_1 \otimes \rho_2 \otimes \rho_3
\end{equation}
and (2) biseparable ($bs$) states, which can be in one of the forms:
\begin{equation}
\begin{multlined}
	\rho^{(1,2,3)}_{bs} =  \rho^{(1)} \otimes \rho^{(2,3)}_{ent}, \\ 
	\rho^{(2,1,3)}_{bs} =  \rho^{(2)} \otimes \rho^{(1,3)}_{ent}, \\
	\rho^{(3,1,2)}_{bs} =  \rho^{(3)} \otimes \rho^{(1,2)}_{ent}
\end{multlined}	
\label{eq:bisep_states}
\end{equation}
where any 2 particles are entangled, but separable from the third.  

All the pure states lie on the boundary between \textit{W} and biseparable states \cite{GUHNE20091}; this indicates that the derived witness is largely determined by the entangled and pure biseparable state training data. The relationship between the purity of the (bi)separable state training data and its distance from the boundary defined by the derived witness is explored more formally in the Appendix.

Fully separable states are constructed by taking the tensor product of 3 single-qubit Haar-distributed states. Likewise, biseparable states may be obtained by taking the tensor product of single- and two-qubit Haar-distributed states, followed by random SWAP operations on each of the three qubits. The last step ensures that all possible permutations shown in \eqref{eq:bisep_states} are considered.

Instead of explicitly measuring the biseparability of each state used for training, we rely on the fact that two-qubit Haar-distributed states are concentrated towards highly entangled states when considering the ensemble's average concurrence \cite{doi:10.1080/09500340110120914}. This result allows us to guarantee that many states labelled as biseparable can only be decomposed into two parts. Consequently, the method will also result in the mislabelling of a small portion of separable states being considered biseparable. However, this does not impact the efficacy of our derived witnesses: a classifier between a set of entangled states and the union of the separable and biseparable states.

As for the entangled state, we again use the Werner state (Eqn. \ref{eq:werner} ) to generate the training data for that class of states. Werner-states are uniformly sampled over the range $p\in[0,0.3)$. 
The training data  consists of $10^5$ 
biseparable and separable states (in equal proportion) and another $10^5$ entangled states.  

\begin{figure}[h!]
	\centering
	\includegraphics[width=7cm]{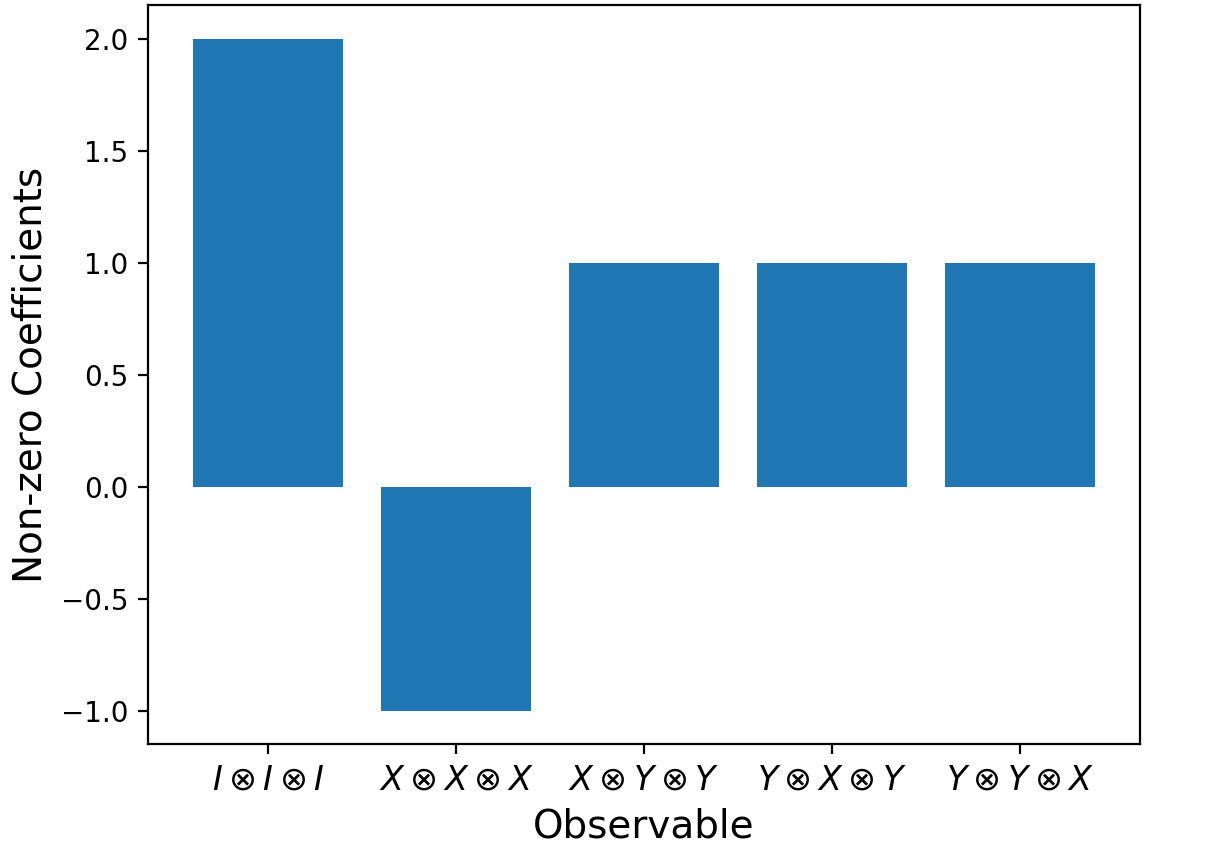}
	\begin{equation}
	    \hat{W}_{GHZ} = 2I\otimes I \otimes I - X\otimes X \otimes X + \text{Per}\{X\otimes Y \otimes Y\}
	    \label{eq:ghz_witness}
	\end{equation}
	\caption{Barplot showing the normalized non-zero coefficients (exceeding 0.01) of the SVM-derived witnesses for 
	the GHZ state. The witness is explicitly defined in \eqref{eq:ghz_witness} below the plot; the operator $Per\{\cdot\}$ denotes the sum of all possible permutations of the operators within the brackets.
	\label{fig:W_GHZ}
	}
\end{figure}

\begin{figure}[]
	\centering
	\includegraphics[width=8.6cm]{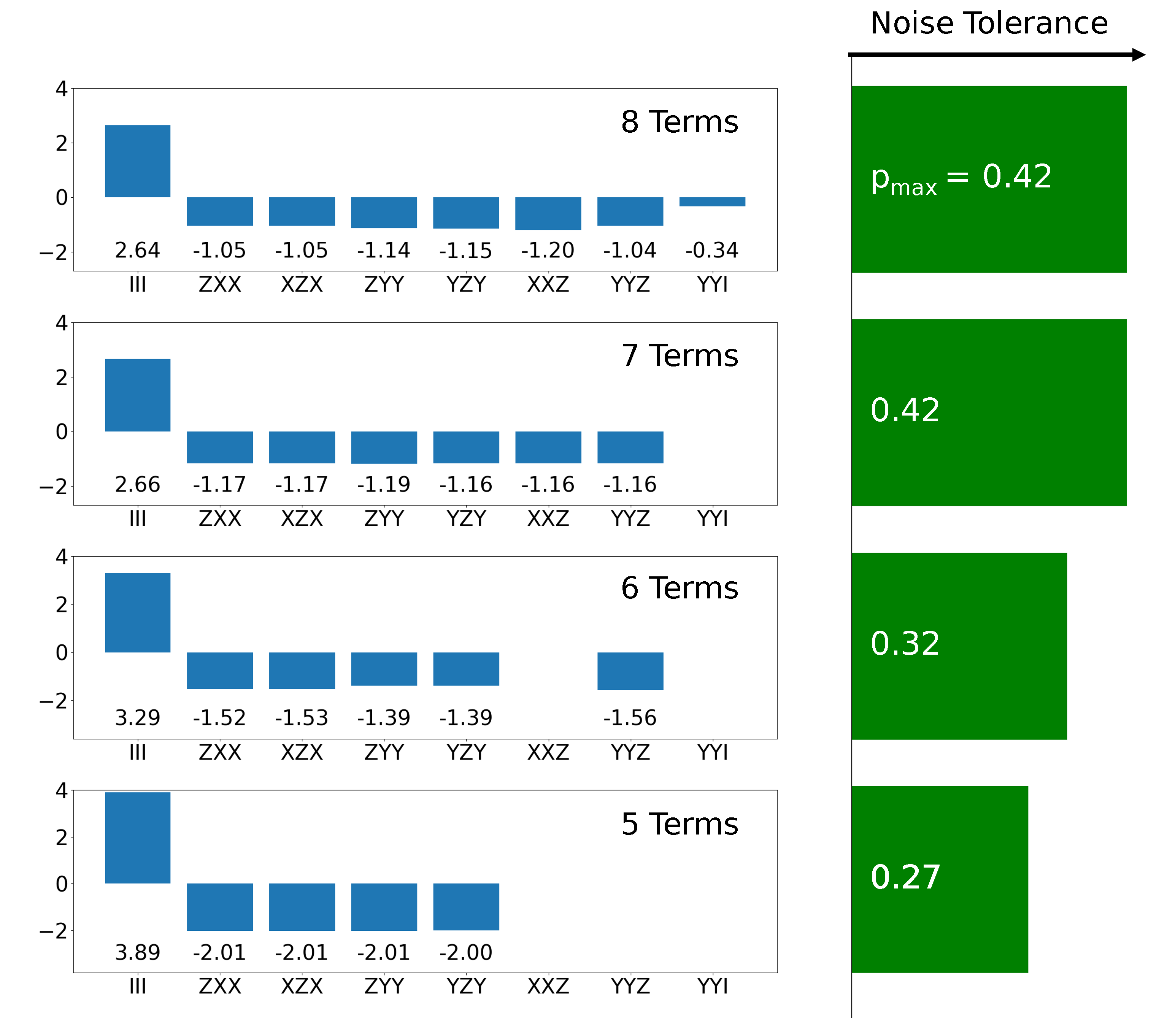}
	\caption{
Systematic removal of features from a witness using Recursive Feature Elimination (RFE).  
The $W$ state witness derived by the SVM initially has 8 coefficients, including the intercept term ($I\otimes I\otimes I$). This initial witness has a noise tolerance of up to 0.42.  
In each iteration of the RFE algorithm we discard the feature with the smallest effect on $p_{max}$. We find that the minimum number of coefficients that the witness can have without misclassifying separable states is 5. However, this reduction decreases the noise tolerance $p\leq$ 0.27). 
\label{fig:RFE}	
}
\end{figure}


Bar graphs indicating the non-zero coefficients for the two derived witnesses (Eqn. \ref{eq:GHZ_W}) are shown in Fig. \ref{fig:W_GHZ} and Fig. \ref{fig:RFE}.  
The coefficients are normalized to a maximum value of 1; any coefficients with a magnitude of less than 0.01 is removed.
We note that the SVM-derived witness for the GHZ state is equivalent to the theoretical
witness obtained with Mermin’s inequality, with the features corresponding to the stabilizers of the GHZ state \cite{toth2005entanglement_stabilizer}. This agreement further validates the SVM scheme. 

What is noteworthy, though, is that the $W$ state is a non-stabilizer state. For such a state, it is not possible to find stabilizing operators that are the tensor products of single-qubit operations \cite{toth2005entanglement_stabilizer, zhan2020detecting__Unfaithful}. Therefore, \textit{W} state witnesses derived in previous works either consisted of non-local stabilizing operators \cite{toth2005entanglement_stabilizer} or were based on the fidelity method \cite{bourennane2004experimental}. 
Our proposed $W$ state witness provides an alternative to existing methods while only relying on local operations.  It tolerates white noise up to $p = 0.42$ (Fig. \ref{fig:RFE}), which is larger than the limit of $p < 8 / 21$ allowed by the fidelity method \cite{bourennane2004experimental}. It should be noted that the noise tolerance of our proposed witness is still within the theoretical limit of $p \approx 0.52$, at which point the state is considered to be biseparable \cite{PhysRevLett.106.190502}.

The derived witnesses can be further simplified through the process of the Recursive Feature Elimination (RFE) algorithm \cite{scikit-learn}. This allows us to systematically reduce the number of features in the witness to lower experimental complexity, but at the expense of lower noise tolerance. 
The goal of RFE is to eliminate less essential features by recursively considering smaller and smaller subsets of the original features using a greedy algorithm. 
Initially, RFE takes the SVM we trained and ranks the features according to their effect on the noise tolerance $p_{max}$, removing the one with the smallest effect; then the model is retrained with the remaining features. 
This process is repeated until the desired number of features is
reached, or when $p_{max}$ drops below a certain threshold. 
Figure \ref{fig:RFE} shows the noise tolerance $p_{max}$ as the RFE algorithm removes more and more terms.

\section{Bipartite Qudits}

In the case of particles residing in higher dimensional ($d>2$) Hilbert spaces, the ansatz witness (Eqn \ref{eqn:W_expect}) would need to be constructed with generalized Pauli matrices.  However, these matrices exist only for dimensions that are prime $P$ (or powers of prime $P^d$); the eigenvectors of the matrices correspond to mutually-unbiased bases (MUBs), while the eigenvalues are complex.  

\begin{figure}
	\includegraphics[width=8.5cm]{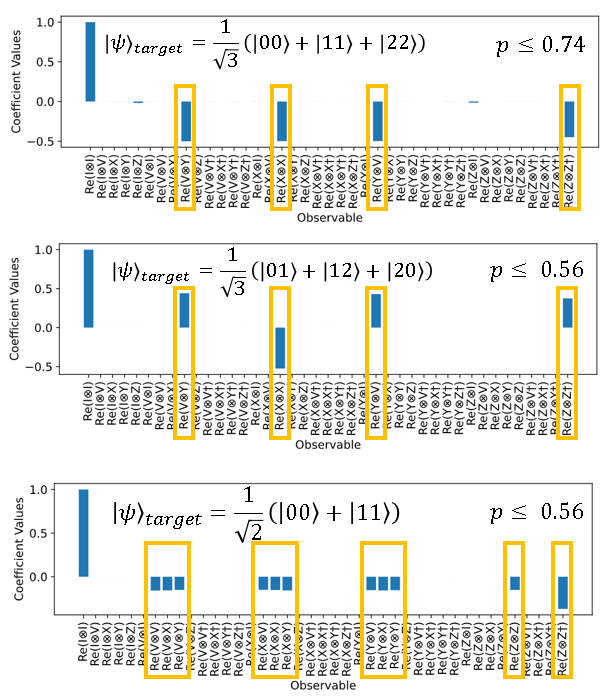}
	\caption{
	The SVM-derived coefficients of the witnesses for three target qudit (d=3) bipartite entangled states are shown.  
	The features with coefficients whose magnitudes exceed than 0.01 are highlighted, and the tolerable white noise for each witness (state) is given on the right hand side.  
	\label{fig:d=3}
	}
\end{figure}

In general, there are $d+1$ MUBs \cite{bandyopadhyay2002new} when $d$ is prime.  The set of corresponding Pauli matrices are:
\begin{equation}
	S = \{ X, Z,  XZ, XZ^2, \cdots, XZ^{d-1}\}, 
\end{equation}
where $X$ is the row-shifted $d$-dimensional identity matrix:
\begin{equation}
X = 
   \begin{pmatrix} 
	0 & 		0 & 		\cdots & 		0 & 		1  \\
	1 & 		0 & 		\cdots & 		0 & 		0  \\
	\vdots & 	\vdots & 	\ddots & 		\vdots & 	\vdots \\
	0 & 		0 & 		\cdots & 		1 & 		0  \\
   \end{pmatrix} 
   \label{eq:Xgenpauli}
\end{equation}
and $Z$ is the diagonal matrix: 
\begin{equation}
Z = 
   \begin{pmatrix} 
	1 & 		0 & 				\cdots & 		0 & 		0  \\
	0 & 		\omega & 		\cdots & 		0 & 		0  \\
   \vdots & 	\vdots & 			\ddots & 		\vdots & \vdots \\
	0 & 		0 & 	\cdots & 		\omega^{d-2} & 	0 \\
	0 & 		0 & 		\cdots & 		0 & 		\omega^{d-1} \\
   \end{pmatrix},\  \omega \equiv e^{j\frac{2\pi}{d}}.
   \label{eq:Zgenpauli}
\end{equation}

From Eq \ref{eq:Zgenpauli}, we see  that the eigenvalues of these generalized Pauli matrices are complex.  
Limiting the dimensionality to $d=3$, the set of generalized Pauli matrices for each particle becomes:
\begin{equation}
	\sigma_1, \sigma_2 \in \{I,V,X,Y,Z,V^\dagger, X^\dagger, Y^\dagger, Z^\dagger\},
\nonumber
\end{equation}
with $Y \equiv XZ$ and $V \equiv XZ^2$.  

The ansatz for the witness given in Eqn \ref{eqn:W_expect} is modified so that it will give real expectation values \cite{Sciara_PhysRevLett.122.120501}; the modification involves hermitianizing each term in the summation:
\begin{equation}
		\hat{W} = \sum_{k_1, k_2}  a_{k_1, k_2} 
	 \left(  \sigma_{k_1} \otimes \sigma_{k_2}  +  \sigma_{k_1}^\dagger \otimes \sigma_{k_2}^\dagger \right).
\end{equation}

Once again, to generate the training data, the separable states are constructed in a Haar-distributed manner, 
and the entangled states are generated by adding white noise to the target pure state (Eqn. \ref{eq:werner}).  
The results for three different states are shown in the form of a histogram in Fig \ref{fig:d=3}; witnesses for each target state are trained with a maximum white noise content of $p = 0.125$.  Even then, the derived witnesses can tolerate significantly more white noise than is present in the training data.  

\begin{figure}
	\includegraphics[width=8.5cm]{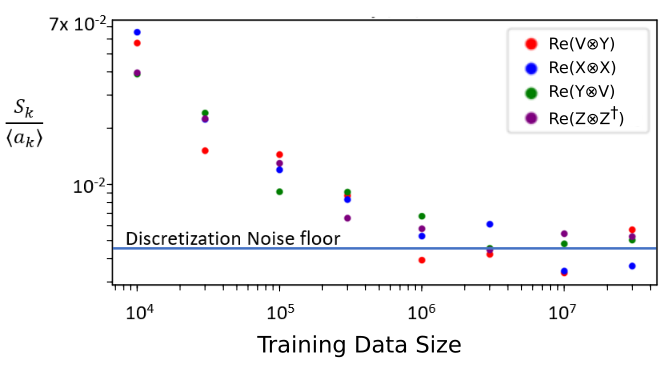}
	\caption{
Convergence testing: How big must the training data be for us to be able to generate a witness whose coefficients converge to its asymptotic value?  The target state
$|\psi\rangle  = \frac{1}{\sqrt{3}}\left( |00\rangle + |11\rangle +|22\rangle \right)$ is used; its coefficients are shown graphically in Fig. \ref{fig:d=3}.  The metric for stability is the ratio of the standard deviation of each non-zero coefficient $S_{a_k}$ to its mean value coefficient $\langle a_k \rangle$.
The training data size is varied from $10^4$ to $3\times{10^7}$.
For large data sizes ($>10^6$), we see asymptotic convergence to a noise floor which we attribute to discretization noise.  
	\label{fig:converge}
	}
\end{figure}

We perform convergence testing to investigate how much training data is required for the SVM-derived witness to converge (Fig \ref{fig:converge}). 
We vary the training data size from $10^4$ to $3\times 10^7$.  For each training data size, a witness for the target state 
$|\psi\rangle  = \frac{1}{\sqrt{3}}\left( |00\rangle + |11\rangle +|22\rangle \right)$
is trained five times, each with different randomly-generated data.  The ratio of the standard deviation of each non-zero coefficient $S_{a_k}$ to its mean value coefficient $\langle a_k \rangle$ is plotted.  
What we observe from Fig. \ref{fig:converge} is that the coefficients have negligible uncertainty (within 1$\%$) and converge to the asymptotic value when training data consisting of more than $10^6$ states points are used.  

In what follows, we shall estimate a simple scaling rule for the number of separable state data points required until the SVM to converge to a single solution. Such a rule is unnecessary for determining the required amount of entangled state data since these states are sampled uniformly according to a single parameter. The question of sampling pure separable states is far more complex due to the larger space of parameters that must be sampled. A pure separable state, for which $d > 2$, is defined by $4(d-1)$ independent parameters. If one were to uniformly sample each parameter, such that each parameter has a probability of $\frac{1}{m}$ of being drawn, this means that we would need $\sim m^{4(d-1)}$ data points to span our space of separable states. This rule implies that if one were to cast $m = 10$, they would need to define $10^8$ data points for $d = 3$. However, this theoretical prediction is in contrast to the empirical results shown in Fig. \ref{fig:converge}, that point to convergence for less than $\sim 10^6$ data points. This gives us confidence that we can efficiently extend our SVM-based method to higher dimensions and more particles.

We do not prescribe a similar scaling rule for the training data required for multipartite entanglement detection. Much like the two-qubit case, the entangled state data of a multipartite system would only depend on a single parameter $p$ for defining a Werner state. Unlike the two-qudit states, the biseparable state training data will have the most significant impact on the derived classifier. A more thorough examination of this result can be found within the Appendix. The proposed approach applied to the scalability of separable state data, cannot be extended to biseparable states due to the difficulty of prescribing a parameterization that spans the entire class of pure biseparable states. The optimal method of sampling biseparable states so as to minimize the number of required data points remains an open question and shall be left to future work.    

%

%
%
\begin{figure*}[h!]
    \includegraphics[width=16cm]{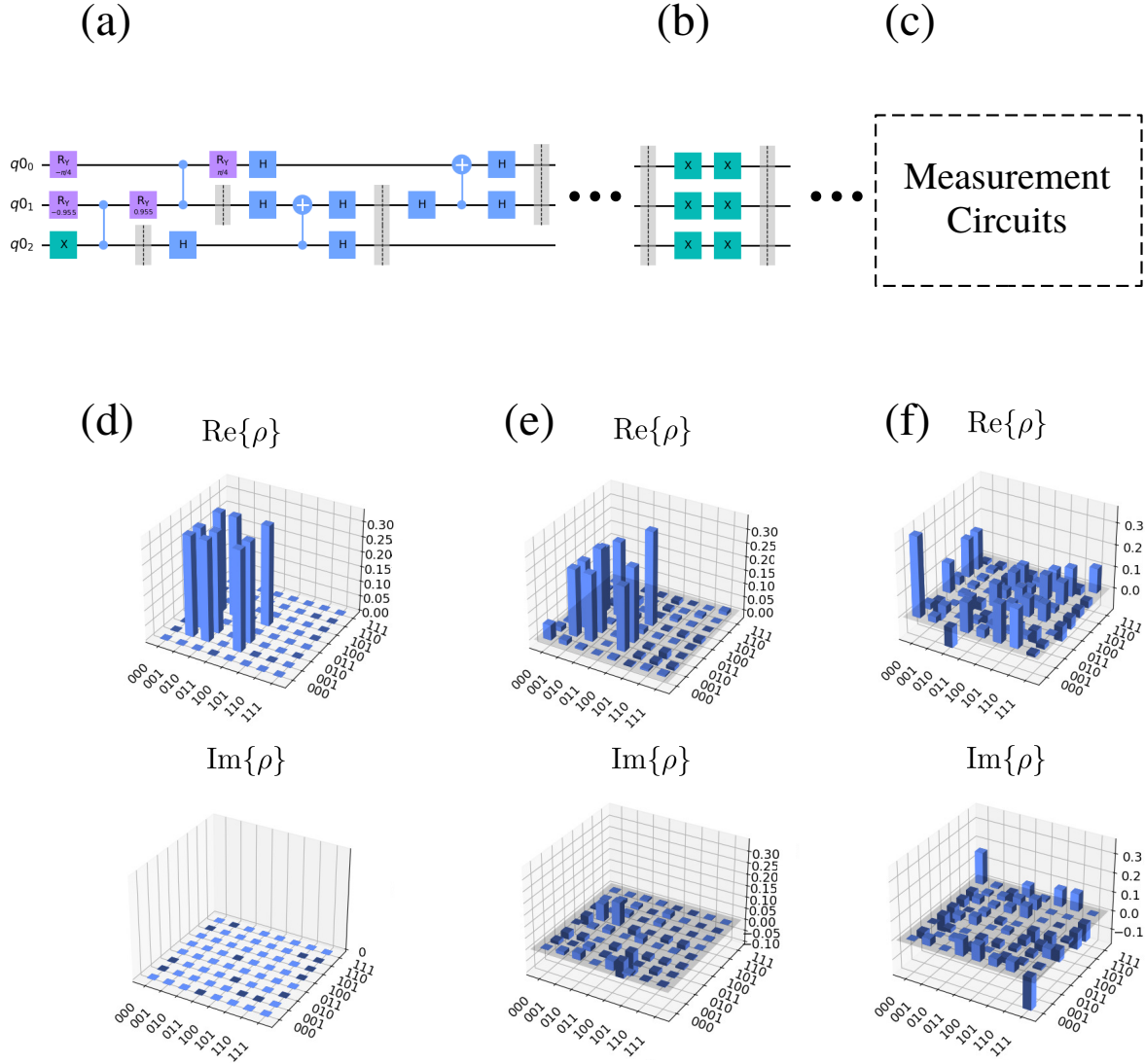}
    \caption{(a) The circuit used to generate a W-state on IBMQ hardware \cite{diker_w_states}. Indeed, this can be done with a series of Z-rotations, whose angles are labelled, Hadamard gates, controlled-NOT operations, and controlled Z-gate operations. (b) The segment in which a variable number of identity operations are appended to the W-state circuit in (a). (c) The measurement circuits corresponding to fidelity and SVM-derived witnesses. (d) The density matrix for an ideal W-state. (e) The density matrix for the W-state generated by (a). (f) The density matrix for a W-state after it has undergone 22 identity operations on \textit{ibm\_perth}. The fidelity values of the states in (e) and (f) to the ideal \textit{W} state are $0.801$ and $0.074$ respectively.}
     \label{fig:circuits_n_dms}
\end{figure*}

Higher $d$-dimensional states will require even larger training data sets; we will need to implement \emph{online} training of the SVM, which involves training the SVM in batches, with different data fed in at each batch.  
This is in contrast to \emph{offline} data, where all the data is available to the SVM for training initially.  
To do this, we implement an SVM structure in Tensorflow \cite{tensorflow2015-whitepaper}, using a single input layer (the features), and single output with no hidden layer. The loss function (Eqn \ref{eqn:lossfunc}) remains the same, but we instead use Adaptive Momentum (ADAM) \cite{kingma2014adam} as the optimizer.  



\section{Physical Verification of \textit{W} State Witnesses}

In this section we seek to quantify the performance of our \textit{W} state witnesses in the face of real non-idealities encountered on noisy-intermediate scale quantum (NISQ) devices such as noise and cross-talk \cite{wood_ieee}. The noise tolerance of each witness is compared against the fidelity method witness \cite{bourennane2004experimental} as a benchmark. The quantum computing platform provided by IBM (IBMQ) was chosen for its ease of access and the maturity of its software development kit, Qiskit \cite{Qiskit}. All physical experiments that follow are performed on the 7-qubit superconducting machine, \textit{ibm\_perth} \cite{lanl_benchmark}. 

We concern ourselves with our witnesses of three qubit \textit{W} states due to the many obstacles associated with creating entangled qudits on NISQ devices. As mentioned prior, there is also a scarcity in \textit{W} state witnesses reliant entirely on local operations in existing literature; this can be attributed to the fact that they are non-stabilizer states. We do not attempt to measure the witness on separable or biseparable states due to the large search space that they occupy.

While bipartite Werner states have been demonstrated on IBMQ by Gårding et al. \cite{garding_mdpi}, 
there does not exist any publicly available implementations of tripartite Werner states. Furthermore, it was found that existing bipartite Werner states implemented on IBMQ show highly unpredictable behaviour as a result of the non-idealities presented in \cite{wood_ieee}. 
Due to the difficulty of generating precise tripartite Werner states on IBMQ, we instead generate a W-state followed by a variable number of identity operations. Fig. \ref{fig:circuits_n_dms}(a), (b), and (c) show the full circuit used to test the separability of tripartite $W$ states followed by a variable number of identity operations on IBMQ hardware. Our means of generating these $W$ states, shown in Fig. \ref{fig:circuits_n_dms}(a), is adopted from a deterministic method of generating arbitrary N-partite $W$ states by Diker \cite{diker_w_states}. We observe in  Fig. \ref{fig:circuits_n_dms}(d) through (f) that the fidelity of the state  to an ideal W-state decreases monotonically with the number of identity operations shown in Fig. \ref{fig:circuits_n_dms}(b). This is based on the fact that these gates are prone to unpredictable sources of error. We use this feature to generate states with controllable values of fidelity to the ideal \textit{W} state. 

After passing the W-state through a selected number of identity operations, we perform projective measurements onto each of the terms of our witnesses. Qiskit allows users to convert a user-defined observable to a series of measurement operators and their respective circuits \cite{Qiskit}. Since our witness consists of operators in a non-diagonal basis, we have an additional step of converting these to a series of Hadamard and Z-rotations. This allows us to approximate the expectations of observables composed of non-diagonal operators when we are limited to only measuring in the Z-basis. The expectation of each witness is calculated from the results of 1024 repeated projective measurements made on each term shown in Fig. \ref{fig:RFE}. Fig. \ref{fig:circuits_n_dms}(c) represents the stage in which the desired witnesses are measured according to the aforementioned technique. Although not shown explicitly, each term of the SVM-derived and fidelity witnesses has its own distinct measurement circuit. This means that an $n$ term witness will require a total of $n\times 1024$ measurements to determine the separability of a given state.  
%
%
%
%
%

\begin{figure}
    \centering
    \includegraphics[width=8.5cm]{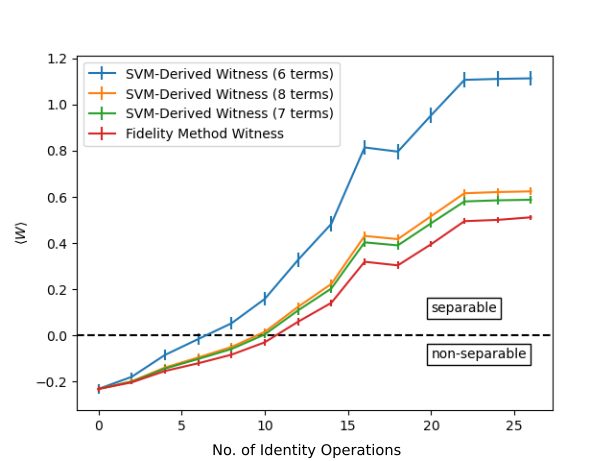}
    \caption{A plot of the number of identity operations versus the expectation of various entanglement witnesses discussed in this paper. Error bars (1-sigma) were calculated by repeated sampling of measurement circuits implemented on IBMQ, and fitting a normal distribution to the measured values. The SVM-derived witnesses have been scaled by a constant factor for better comparison with the fidelity-method witness.}
    \label{fig:id_v_w_ibmq}
\end{figure}

Fig. \ref{fig:id_v_w_ibmq} shows that our witnesses with 6, 7, and 8 terms tolerate an amount of noise comparable to the fidelity method witness before detecting separability \footnote{One may notice that all witnesses experience ``jumps''   in their expectation values at regular intervals. This is attributed to changes in the operational environment such as regular re-calibration performed on devices.}. The SVM-derived witnesses' advantage lies in the fact that they require far less measurements than the 20 term fidelity method witness when measured in the Pauli basis \cite{bourennane2004experimental}.

To supplement these physical experiments, we perform a series of numerical simulations of the following state for $p$ ranging from $0$ to $1$: 

\begin{equation}
    \rho = (1-p)\ket{W}\bra{W} + \frac{p}{8}I.
    \label{eq:w_werner}
\end{equation}

\begin{figure}[]
    \centering
    \includegraphics[width=8 cm]{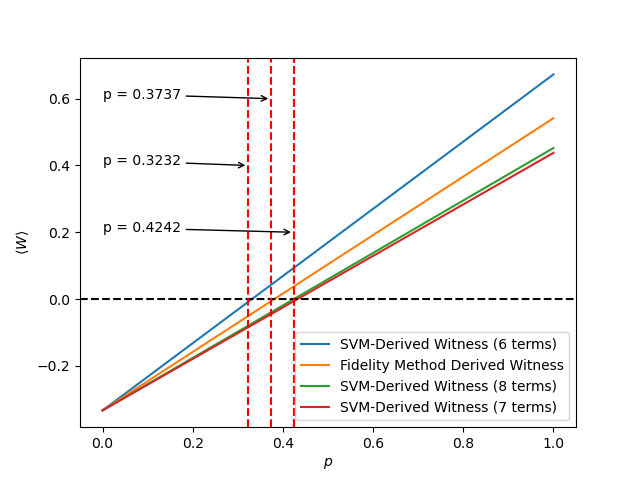}
    \caption{A comparison between the noise tolerance of various SVM-derived witnesses and the Fidelity method witness. Simulations are based on the parameterized W-Werner state defined in (\ref{eq:werner}). Zero-crossings are indicated with dashed red lines.}
    \label{fig:id_v_w_sim}
\end{figure}

The simulation results shown in Fig. \ref{fig:id_v_w_sim} \emph{agree with} our experimental data in the sense that the 6, 7, and 8 term witnesses all show comparable noise tolerance to the fidelity method witness. It is important to note that although the 7- and 8-term SVM-derived witnesses have higher noise tolerance than the fidelity method, the physical experiments in Fig. \ref{fig:id_v_w_ibmq} show that this is not necessarily the case in practice. We attribute this result to the fact that the noisy \textit{W} states that we generate do not resemble the states used for training presented in \eqref{eq:w_werner}. 

\section{Discussion and Conclusion}
While in this work we have derived entanglement witness using the SVM method, we would like to remark the important differences between an SVM classifier and an entanglement witness.

First, an entanglement witness is only partially a classifier, since it is only required to place all separable states on one side of the hyperplane while not required to (as well as not possible to) place all entangled states on the other side. In contrast, an ideal classifier should strictly separate the two classes of states. Therefore, it is important when generating training data for the entangled states to ensure that the entangled states introduced in the training should still make it possible to have a hyperplane separating the entangled states from the separable states. As a result of the Hahn-Banach theorem, a linear SVM will necessarily return a hyperplane between two classes of data so long as they form two disjoint convex sets \cite{boyd_optimization}. Thus, it is important to select the entangled states from a convex set so that they may be distinguished from the separable states, which are known to be closed and convex \cite{GUHNE20091}. The set of Werner states is convex when $p$ is defined over a continuous interval, which implies that the entangled states used throughout this work are sampled from convex sets. Indeed, this motivates our decision to use the Werner states for entangled state data generation throughout this work.  

Second, the SVM classifier tries to maximize the distance between the boundaries of the two classes of states, whereas an ideal entanglement witness should be a hyperplane as close as possible to the convex space formed by the separable states. This difference again points to the importance of generating the training data for the separable states as close to the boundary as possible on one hand, while on the other hand, having training data for the entangled states as close to the boundary as possible while still maintain convexity.

Thirdly, as suggested in Fig.1(c), the training data for the separable states can sampled from a much smaller subset of the set of all separable states, since only states at the boundary closest to the target states, i.e. the support vectors, have a significant influence on the derived witness. As detailed in the Appendix, these correspond to pure biseparable states for the case of the tripartite qubit states.

These differences not only elucidate the limitation of this ML approach in deriving entanglement witnesses, but also help to further improve the SVM approach by improving upon the training data.

In summary, we have demonstrated using linear support vector machines (SVM) to 
derive entanglement witnesses for both bipartite and tripartite qubit systems, as well as bipartite qutrit systems.  
This method generates witnesses that require only local measurements and whose tolerance to noise is on par with what is currently found in the literature.  
In addition to deriving witnesses for stabilizer states, our method also yields witnesses to non-stabilizer states.  
Furthermore, we demonstrate the ability to systematically reduce the number of features in the witness, allowing us to reduce experimental complexity at the expense of noise tolerance. Finally, we have verified the SVM-derived witness for the W-state on physical IBMQ hardware, with realistic and uncontrolled noises added by quantum gates, and have shown that the SVM-derived witnesses has a comparable noise tolerance to the fidelity witness while requiring significantly fewer number of measurements.
This SVM method is straightforward to extend to higher-dimensions $d$ for bipartite systems, involving online learning, and may also be computationally efficient for deriving the witnesses of systems consisting of larger particle numbers $N$. The fact that the number of measurement terms of the SVM-derived witnesses can be decreased at the cost of noise tolerance makes them an appealing benchmark of multi-partite and higher-dimensional entanglement in modern NISQ devices.

\section{Acknowledgments}
This work was funded by NSERC (National Sciences and Engineering Research Council of Canada) and the US Army 
Research Laboratory.  We thank IBM for free access to their IBMQ quantum computing systems. We would also like to thank Sanjaya Lohani and Thomas Searles for valuable discussion pertaining to the IBM Q systems. Finally, we thank Marcus Huber for our insightful discussion on multipartite entanglement measures.  

\section*{Appendix: Sampling Random Separable and Biseparable States}

By virtue of being derived via statistical learning methods, the accuracy of our witnesses depends heavily on having representative training data. We found that our means of sampling random separable and biseparable states used for training the SVM corresponding to our \textit{W} state witnesses had a dramatic impact on the witnesses' accuracy in Section IV. 

When choosing a data set, it is important to note that data points closest to the boundary between classes will have the most profound effect on the hyperplane corresponding to a derived witness. We shall begin our exploration by searching for a correspondence between the different methods of sampling data and their concentration around the boundary defined by our classifier. Determining these data points, referred to as \textit{Support Vectors}, could allow us to considerably decrease the size of a data set by primarily considering points close to the defined boundary \cite{statistical_learning_text}. However, given the complex geometric properties of sets in Hilbert space, it is important to quantify the role of points interior to the set of biseparable states in training our SVM. In what follows, we justify our choice of using entirely pure states as (bi)separable training data throughout this work. Particular emphasis is placed on the training data used for 3-qubit \textit{W} states.

For the purposes of this study, pure and mixed fully-separable states are constructed by taking the tensor products of three single-qubit Haar-distributed and Mai-Alquier (MA) distributed states \cite{PhysRevResearch.3.043145} respectively. Indeed, the MA-distributed states $\rho_{MA}$ may be defined as a convex combination of Haar-distributed states $\ket{\psi_j}$ weighted by a Dirichlet random variable $\mathbf{x}$ distributed according to $\text{Dir}(\mathbf{x}|\mathbf{\alpha})$:

\begin{equation}
    \text{Dir}(\mathbf{x}|\mathbf{\alpha}) = \frac{\Gamma(\sum_{j=1}^{K}\alpha_j)}{\prod_{j=1}^{K}\Gamma(\alpha_j)}\prod_{j=1}^{K}x_j^{\alpha_j-1},
\end{equation}

\begin{equation}
    \rho_{MA} = \sum_{j=1}^{K}x_j \ket{\psi_j}\bra{\psi_j}.
\end{equation}

$\Gamma$ represents the Gamma function, $\alpha_j$ is a positive real number referred to as the ``concentration parameter" and $K\geq 2$ is often referred to as the ``number of categories" of our Dirichlet distribution. In the case of MA-distributed states defined in \cite{PhysRevResearch.3.043145}, one can assume that the weights $x_j$ correspond to a symmetric Dirichlet distribution, such that $\alpha_j = \alpha$. Lohani et al. have shown in \cite{PhysRevResearch.3.043145, Lohani_2022} that one may tune the statistics in purity of an ensemble of MA-distributed states by their choice of $\alpha$ and $K$. The single- and two-qubit MA-distributed states used in this section have weights assigned according to a symmetric Dirichlet distribution such that $\alpha_j = \alpha = 0.1$ and $K=4$.

\begin{figure}
    \centering
    \includegraphics[width=7.5cm]{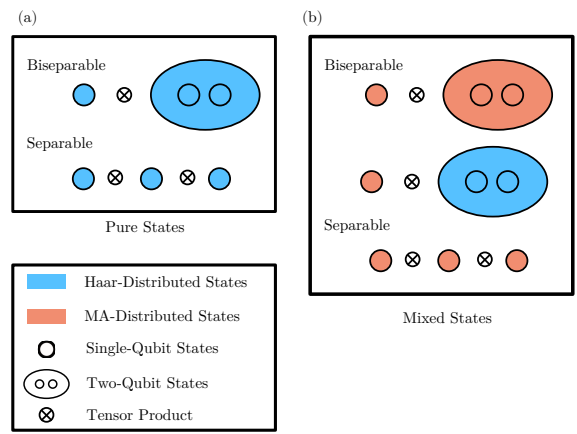}
    \caption{The schemes used to generate random fully separable and biseparable states for training the model presented in Section II. Pure and mixed states are shown in (a) and (b) respectively. We ensure that our training data covers multiple classes of biseparable states by applying random SWAP operations between qubits.}
    \label{fig:sep_bisep_states}
\end{figure}

Pure and mixed biseparable states $\rho_{bs}^{(1,2,3)}$ are sampled by taking the tensor products of single- and two-qubit Haar and MA-distributed states. Much like their fully-separable counterparts, biseparable pure states can be found by taking the tensor product of single- and two-qubit Haar-distributed states. Mixed biseparable states are obtained by taking the tensor product of either: a single-qubit Haar-random state with a two-qubit MA-distributed state, or single- and two-qubit MA-distributed states. The schema for sampling fully-separable and biseparable states is summarized in Fig. \ref{fig:sep_bisep_states}.

\begin{figure}
    \centering
    \includegraphics[width=8cm]{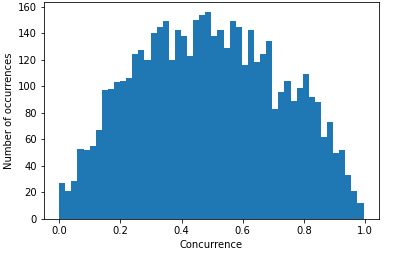}
    \caption{A histogram of concurrence for bipartite MA-distributed states for which $\alpha_K = \alpha = 0.1$ and $K=4$. The data set consists of 5000 sample states.}
    \label{fig:conc_ma_states}
\end{figure}

\begin{figure}
    \centering
    \includegraphics[width=8cm]{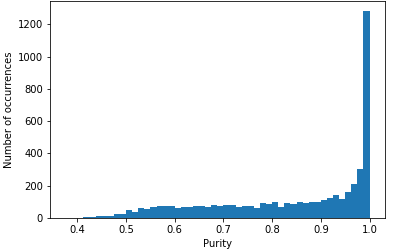}
    \caption{A histogram of purity for bipartite MA-distributed states for which $\alpha_K = \alpha = 0.1$ and $K=4$. The data set consists of 5000 sample states.}
    \label{fig:pur_ma_states}
\end{figure}

\begin{figure}
    \centering
    \includegraphics[width=8cm]{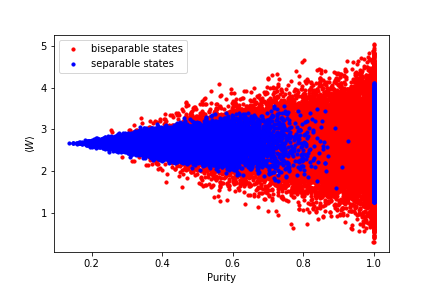}
    \caption{A scatter plot that depicts the relationship between the purity and $\expval{W}$ of our 8-term \textit{W} state witness derived in Section II. 60 000 sample states are obtained by taking the tensor product of single- and two-qubit Haar- and MA-distributed states according to the scheme introduced in Fig. \ref{fig:sep_bisep_states}. All MA-distributed states (single and two-qubit states) are tuned such that $\alpha_K = \alpha = 0.1$ and $K = 4$.}
    \label{fig:pur_v_w}
\end{figure}

\begin{figure}
    \centering
    \includegraphics[width=8cm]{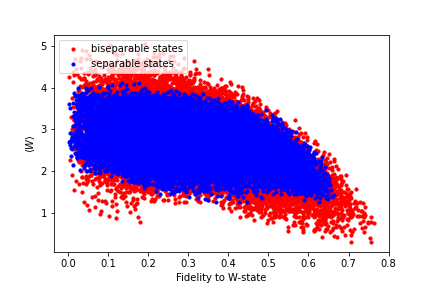}
    \caption{A scatter plot that depicts the relationship between the fidelity, as defined in \cite{mike_n_ike}, to the ideal \textit{W} state and $\expval{W}$ of our 8-term witness derived in Section II. We use a total of 60 000 separable and biseparable states obtained according to the scheme in Fig. \ref{fig:sep_bisep_states}. All MA-distributed states (single and two-qubit states) are tuned such that $\alpha_K = \alpha = 0.1$ and $K = 4$.}
    \label{fig:w_v_fid}
\end{figure}

We find that it is less common to obtain mixed biseparable states from the scheme in Fig. \ref{fig:sep_bisep_states} depending on the average purity of the sampled MA-distributed states; mixed states derived from this scheme are most likely to be fully separable. When setting $\alpha_k = \alpha = 0.1$ and $K= 4$, we find a correspondence between the average purity of random bipartite MA-distributed states, and their respective concurrence. Histograms of the concurrence and purity of bipartite MA-distributed states are depicted in Fig. \ref{fig:conc_ma_states} and \ref{fig:pur_ma_states} respectively. 

After having trained an SVM for a \textit{W} state witness using the aforementioned (bi)separable state training data along with the entangled states described in Section II, we find a direct correspondence between the purity of a state and its distance to the boundary defined by our classifier. This result is shown in Fig. \ref{fig:pur_v_w}. The fact that highly pure states are both furthest and closest to the boundary defined by our classifier is an indicator that our method of sampling likely spans the surface of the biseparable states. Much like pure states, those with high fidelity to the ideal \textit{W} state will be close to this boundary, as shown in Fig. \ref{fig:w_v_fid}.  

The trend that separable states are ``encapsulated" by those that are biseparable in Fig. \ref{fig:conc_ma_states} and \ref{fig:pur_ma_states} resembles the structure presented by Acín and colleagues in \cite{PhysRevLett.87.040401}, an early formulation of the geometric properties of the three-qubit states. Two questions naturally arise from this study: (1) ``Are mixed (bi)separable states necessary in training an SVM for the detection of three-qubit entanglement?" and (2) ``Is fully-separable state training data necessary for deriving a three-qubit witness?". 

The first of these two questions can be verified by comparing the derived witnesses that result from using mixed and pure states sampled according to Fig. \ref{fig:sep_bisep_states}, in which a total set of $2.5\times 10^6$ combined separable and biseparable and $2\times 10^6$ entangled state data points are sampled. For the case of \textit{W} states, we find that the resulting coefficients remain the same as Fig. \ref{fig:RFE}: which involved only pure separable and biseparable states during training. Indeed, this result is unsurprising considering that the pure biseparable states correspond to our support vectors, and have the most significant effect on the classifier's derivation. 

\begin{figure}
    \centering
    \includegraphics[width=10cm]{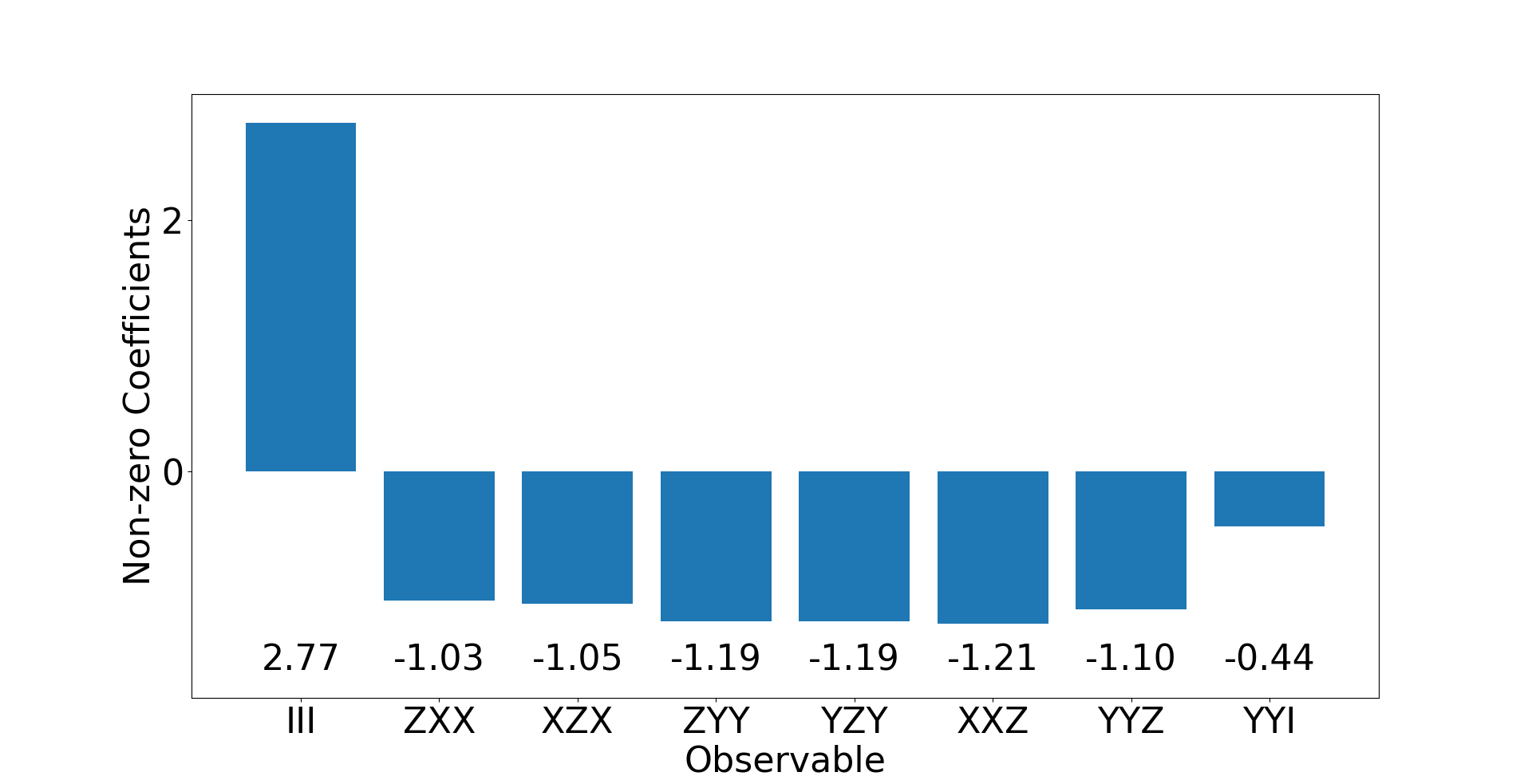}
    \caption{The 8-term \textit{W} state witness derived with $2\times 10^{6}$ entangled and $500\times 10^5$ pure biseparable state data points.}
    \label{fig:w_wit_8_term_no_sep}
\end{figure}

The second question can be answered by discarding all training data from the previous case, with exception to the entangled and pure biseparable state training data. Upon retraining the SVM, we obtain the witness shown in Fig. \ref{fig:w_wit_8_term_no_sep}. The largest discrepancy in the derived coefficients also corresponds to the smallest term, $Y\otimes Y \otimes I$, whose impact on the witness's classification is negligible. And thus, the noise tolerance remains unchanged from the value of p = 0.42 reported in Fig. \ref{fig:RFE}. The fact that one can derive a \textit{W} state witness with only pure biseparable state data indicates the importance of informing one's choice of training data from physical arguments. Future work shall build on this result further by demonstrating how the required training data scales when deriving witnesses of entanglement in larger multiparticle systems.

\newpage
\bibliography{Pubs}

\end{document}